\documentclass{mem}
\usepackage{natbib}\usepackage{txfonts}\usepackage{balance}
\usepackage{graphicx}
\usepackage[a4paper]{hyperref}
\idline{75}{282}
\begin{document}
\def\teff{$T\rm_{eff }$}
\def\kms{$\mathrm {km s}^{-1}$}

\title{
Spectral energy distribution of gamma-ray binaries:
}

   \subtitle{Sources and Processes}

\author{
V.\,Bosch-Ramon\inst{1} 
          }

  \offprints{V. Bosch-Ramon}

\institute{
Dublin Institute for Advanced Studies, 31 Fitzwilliam Place, Dublin 2,
Ireland.
\email{valenti@cp.dias.ie}
}

\authorrunning{Bosch-Ramon}

\titlerunning{Gamma-ray binaries}

\abstract{
Gamma-ray binaries are suitable sources to study
high-energy processes in jets and outflows in general. In the last years,
there has been a lot of activity in the field of gamma-ray binaries
to identify the different factors
that shape their non-thermal spectra, which ranges from
radio to very high energies,
as well as their lightcurves. In this work, I discuss the main
aspects of the non-thermal emission in this class of objects, 
which presently includes high-mass microquasars, high-mass binaries hosting 
a non-accreting pulsar and, probably, massive star binaries; few potential
candidates to be gamma-ray binaries are also presented. Finally,
the importance of gamma-ray absorption is discussed, and the main 
physical ingredients, which are likely involved in the non-thermal radiation in gamma-ray binaries,
are briefly considered.
\keywords{X-ray: binaries -- Gamma-rays: theory -- Radiation mechanism: non-thermal}
}
\maketitle{}

\section{Introduction}

Gamma-ray binaries are binary systems formed by two objects, generally one non-degenerated star and a compact object or
another non-degenerated star, which emit gamma-rays. 

Three gamma-ray binaries have been detected so far in high- (HE: $\sim 0.1-100$~GeV) and very high-energy (VHE: $\sim
0.1-100$~TeV) gamma-rays: LS~5039, LS~I~+61~303 and Cygnus~X-1 (at $\sim 4\,\sigma$) 
\citep{kniffen97,paredes00,aharonian05a,aharonian06a,albert06,albert07,acciari08,abdo09a,abdo09b,sabatini10}. In addition to
these three sources, there are a gamma-ray binary and a candidate for it, PSR~B1259$-$63 and HESS~J0632$+$057, respectively,
detected above 100 GeV, but not below \citep{aharonian05b,aharonian07}. Finally, Cygnus X-3 has been detected in the HE but
not in the VHE band \citep{tavani09a,abdo09c,saito09}, as well as probably $\eta$-Carina, which could have been detected only
in the HE range \citep{tavani09b}. Among all these sources, there are two confirmed high-mass microquasars (HMMQ) (Cygnus~X-1
and Cygnus~X-3), one confirmed high-mass binary hosting a non-accreting pulsar (PSR~B1259$-$63), possibly a massive star
binary ($\eta$-Carina), two high-mass binaries harboring a compact object of unknown nature (LS~5039 and LS~I~+61~303; see,
e.g., \citealt{dubus06a,chernyakova06,romero07,bosch09b}), and a high-mass binary candidate (HESS~J0632$+$057). 

In the following, we briefly introduce the different types of source that have been detected in gamma-rays
(Sect.~\ref{src1}), as well as others that may be detectable in the nearby future (Sect.~\ref{src2}). The importance of
gamma-ray absorption in gamma-ray binaries is also discussed (Sect.~\ref{abs}), and a short inventory is made, together with
some remarks, of all the different physical ingredients that are probably to be considered in these sources
(Sect.~\ref{fi}).  For a very recent observational overview of gamma-ray binaries, we address the reader to \cite{paredes10}.

\section{Physical scenario of established gamma-ray sources}\label{src1}

\subsection{High-mass microquasars}

\begin{figure*} 
\center
\includegraphics[width=8cm]{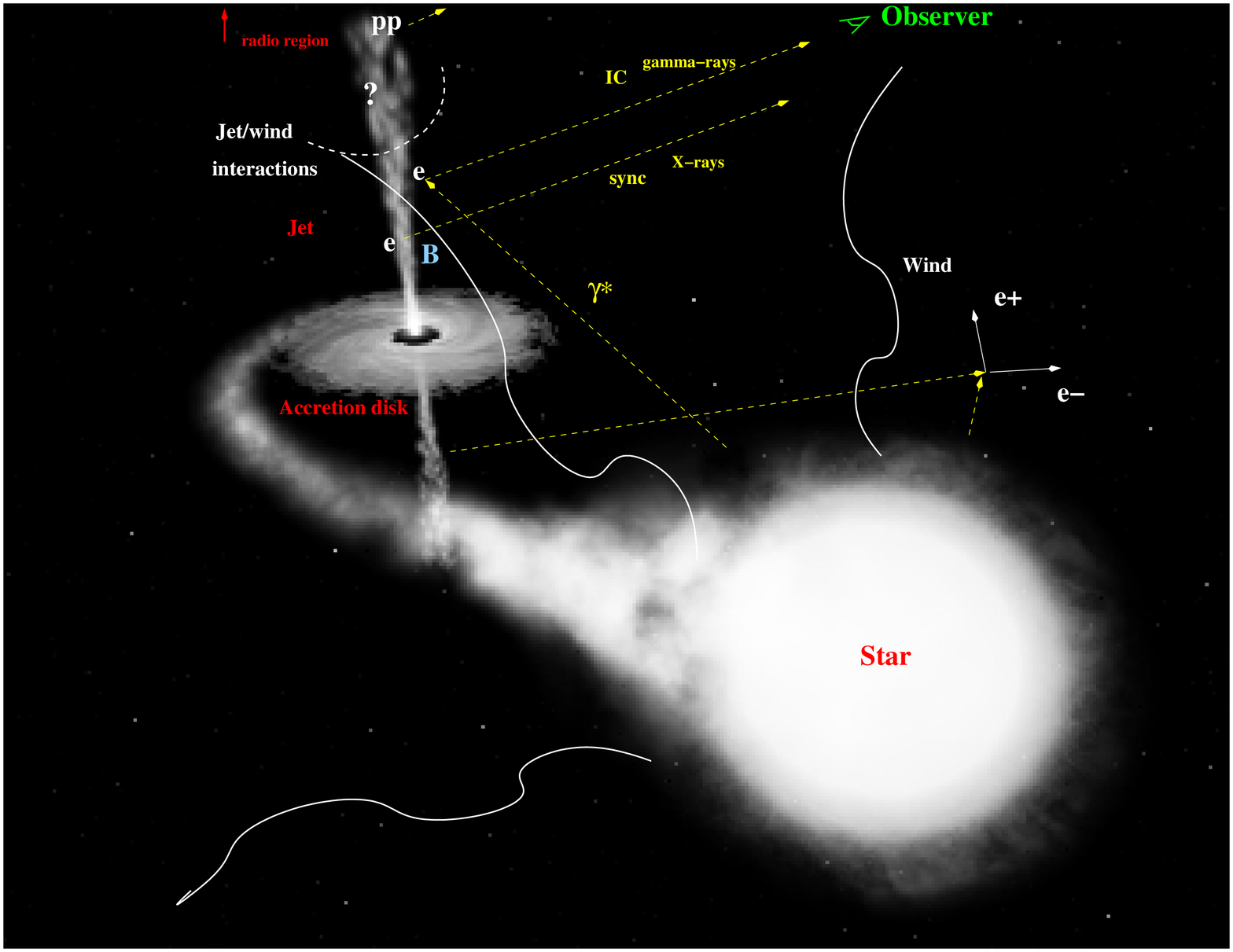}
\caption{\footnotesize
Illustrative picture of the microquasar scenario, in which
the main elements and processes relevant to this work are shown
(background image adapted from ESA, NASA, and F\'elix Mirabel).
} 
\label{mq}
\end{figure*}

Microquasars are X-ray binaries with radio jets \cite[see, e.g.,][]{mirabel99,ribo05}, and when hosting a massive star, they
are called HMMQ. The power supply in these objects can be either accretion or an accreting rotating black-hole, and emission
would be directly powered by a jet launched in the inner regions of the accretion disk. The jet can produce non-thermal
populations of relativistic particles via diffusive shock acceleration or other mechanisms at different spatial scales (for a
recent review, see \citealt{bosch10a}; see also, e.g., \citealt{levinson96,romero03,paredes06,rieger07}). In the context of
HMMQ, the most efficient gamma-ray mechanism would be inverse Compton (IC) scattering of the stellar photons
\citep[e.g.][]{paredes00,dermer06,bosch06a}. A sketch of the microquasar scenario is shown in Fig.~\ref{mq}.

Beside the jet itself, stellar wind-jet dynamical interactions at the binary spatial scales could also lead to non-thermal
emission in HMMQs \citep{perucho08}, and that the likely clumpy nature of the wind \citep{owocki06} could lead to HE and VHE
flares \citep[see, e.g.,][]{araudo09}. 

\begin{figure*} 
\center
\includegraphics[width=11cm]{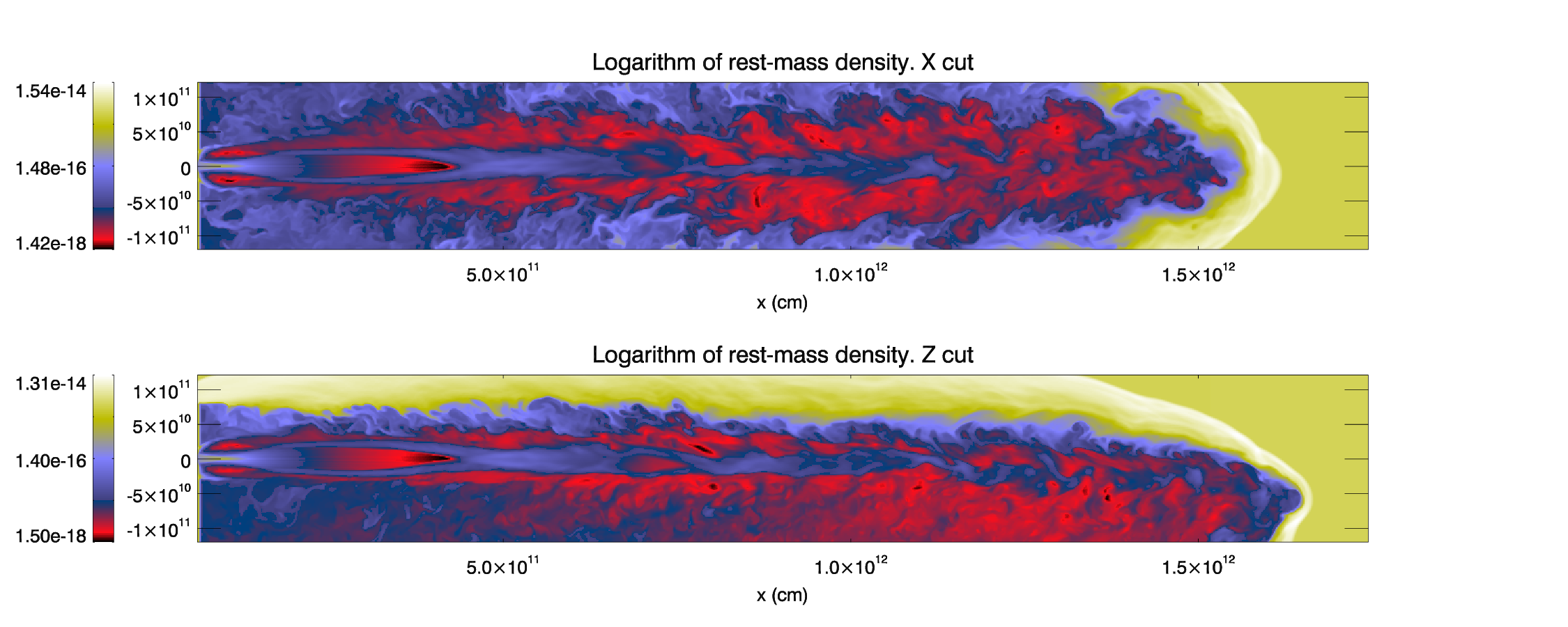}
\caption{\footnotesize
Two-dimensional map of the density of a jet interacting with the stellar wind, which is coming from the top.
} 
\label{jetint}
\end{figure*}

An example of the importance of the wind-jet interaction is shown in Fig.~\ref{jetint},  in which it is shown a 2-dimensional
map of the density of a jet of power $10^{35}$~erg~s$^{-1}$ that is launched and interacts with the stellar wind, which is
coming from the top. Note that the jet is suffering significant bending and strong disruption not far beyond an asymmetric
recollimation shock. Jets significantly more powerful, of up to $10^{37}$~erg~s$^{-1}$, may be still affected by jet
disruption due to the growth of non-linear instabilities triggered in the recollimation shock \citep[see][]{perucho10}.

\subsection{High-mass binaries with a non-accreting pulsar}

\begin{figure} 
\center
\includegraphics[width=6cm]{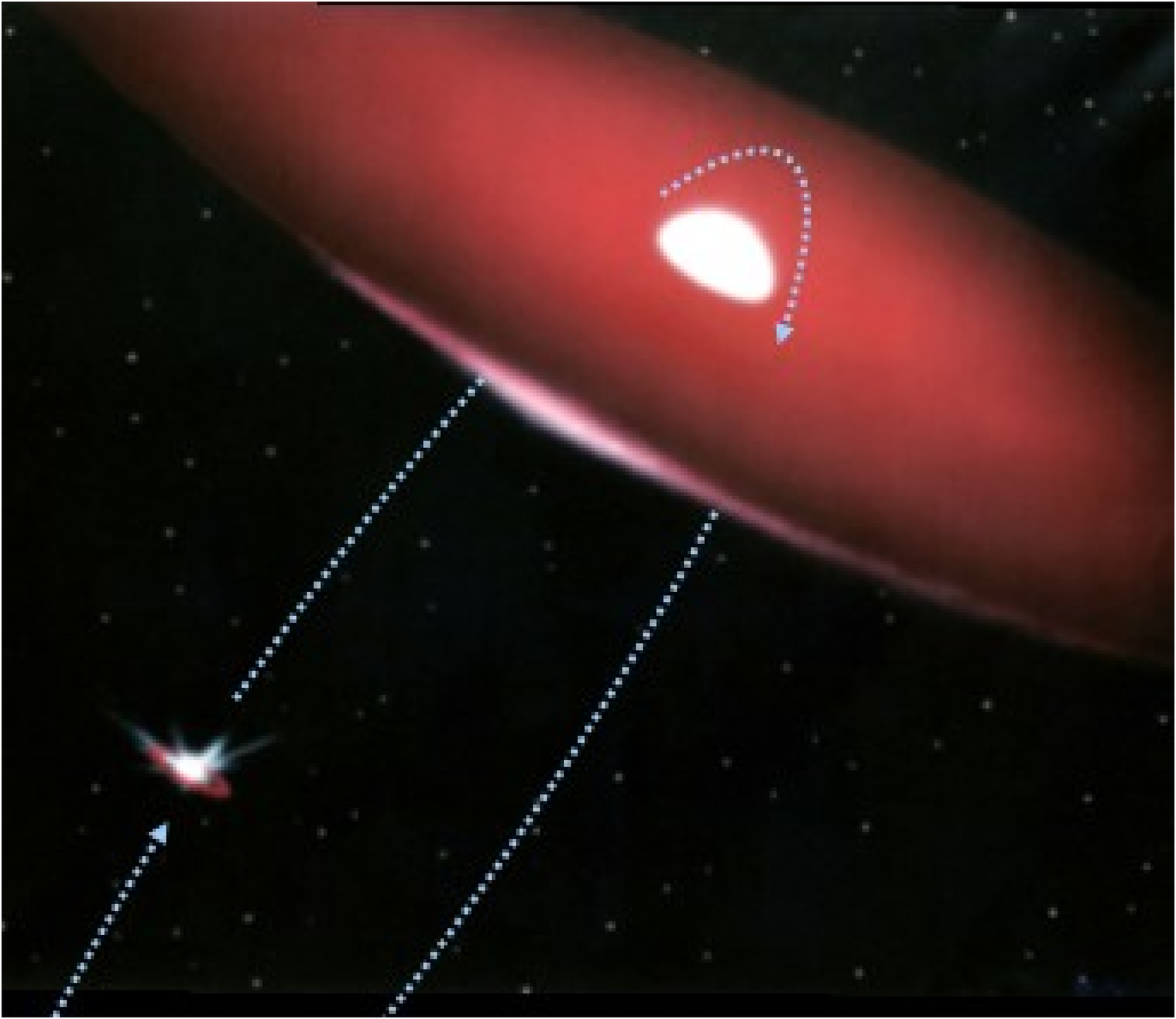}
\caption{\footnotesize
Illustration of a binary system consisting of a pulsar and a Be star. The pulsar orbits a massive Be star with a
disk-like outflow of stellar material (from Hubble archive). Around periastron, the pulsar outflow interaction 
with the stellar
wind is the strongest, leading to particle acceleration and non-thermal emission. More regular and circular systems
may yield stable emission all along the orbit.
} 
\label{pul}
\end{figure}

In young pulsar binary systems, the non-thermal emission is expected to be generated in the region of collision between the
massive star (non-relativistic) and the pulsar (relativistic) winds
\citep[e.g.][]{maraschi81,tavani97,kirk99,dubus06a,khangulyan07,neronov07}, although the shocked pulsar wind, advected away
and accelerated by strong pressure gradients \citep{bogovalov08}, could also be a source of Doppler boosted emission
\citep{khangulyan08a}. The unshocked pulsar wind may be also a source of gamma-rays \citep[e.g.][]{khangulyan07}, which may
suffer reprocessing in the stellar field due to pair creation \citep[see, e.g.,][]{sierpowska07}. This reaccelerated flow may
release kinetic energy far from the shock region, increasing the complexity of the phenomenology of the standard scenario,
since this could affect the non-thermal emission in the whole spectrum. A picture representing the scenario discussed in this
section is shown in Fig.~\ref{pul}.

\subsection{Massive star binaries}

\begin{figure*} 
\center
\includegraphics[width=10cm]{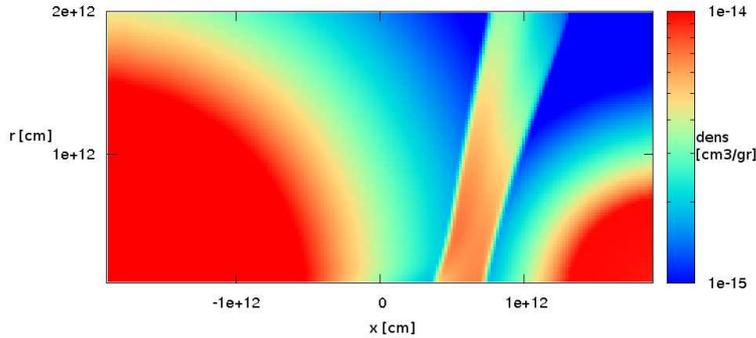}
\caption{\footnotesize
Two dimensional map of the density distribution in a system of two O stars with colliding winds. The wind of the star
to the left is 10 times denser than that of the star to the right.
} 
\label{col}
\end{figure*}

Massive star binaries are systems in which a strong shock takes place between the winds of the stars. These shocks have
speeds of up to few 1000~km~s$^{-1}$ and are collisionless and strongly supersonic, and particle acceleration and non-thermal emission up to
gamma-ray energies, mainly through IC, could be efficient in there
\citep{eichler93,benaglia03,bednarek05,reimer06,debecker07,pittard10}. In Fig.~\ref{col} we show the result of a
2-dimensional simulation of the hydrodynamical interaction between the winds of two O stars. The strong shock between the two
stellar winds is clearly visible in the density jump.

\section{Other potential gamma-ray binaries}\label{src2}

Beside HMMQ, non-accreting pulsar high-mass binaries and massive star binary systems, all of them already detected in
gamma-rays, other binaries are potential candidates to be also found in this energy range. 

\begin{figure} 
\center
\includegraphics[angle=270,width=6.5cm]{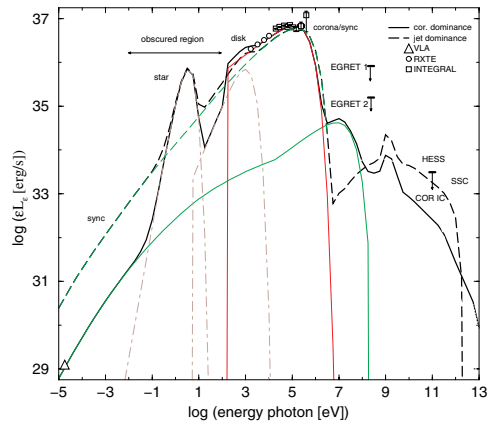}
\caption{\footnotesize
Computed spectral energy distribution of the non-thermal emission from 1E~1740.7$-$2942 for two situations. In one case, 
the hard X-rays come from a corona (long dashed line), whereas in the other, they are of synchrotron origin and come from the
jet. Absorption in the accretion disk and corona photon fields has been taken into account (see the two dips below and above
$\sim$ GeV energies). For details, see Bosch-Ramon et al. (2006b).
} 
\label{lmqs}
\end{figure}

Low-mass microquasars have been also proposed to be HE and VHE emitters. In these objects, the most efficient gamma-ray
mechanisms could be external IC with accretion photons, synchrotron self-Compton or hadronic interactions in the inner-most
regions of the jet base \citep[e.g.][]{atoyan99,bosch06a,romero09}. In microquasars in general, the corona (or the base of
the jet) could be also a non-thermal emitting region (e.g. \citealt{gierlinski99}; see also \citealt{romero10}), as well as
the termination region of the jet in the ISM \citep{bordas09}. In Fig.~\ref{lmqs}, we show the computed spectral energy
distribution for a powerful jet of a low-mass microquasar. The dominant IC component is either IC scattering off corona
photons or synchrotron self-Compton; note the soft spectrum above GeV energies due to the Klein Nishina effect. Gamma-ray
absorption in the accretion disk and corona fields can be significant in this context (for details, see \citealt{bosch06b}).

Accreting pulsar X-ray binaries might be also sources of gamma-rays, as it has been suggested by, e.g., \cite{romero01} and
\cite{sguera09}, each of these works pointing to different powering mechanisms: proton beams moving along the pulsar
magnetospheric magnetic lines and colliding with the accretion disk, or sudden ejections of matter through the formation of a
magnetic tower preceded by high-rate accretion events. This high-rate accretion events would lead to overcome the magnetic
pressure of the pulsar magnetosphere allowing for jet formation \citep[see, e.g.,][and references therein]{massi08}.

\section{Impact of gamma-ray absorption}\label{abs}

Other non-thermal emitting regions different from those already discussed could exist in any type of compact high-mass
binary. If gamma-ray absorption takes place under the photon field of the star
\citep{ford84,protheroe87,moskalenko94,boettcher05,dubus06b,khangulyan08b,reynoso08}, the whole binary system could be an
efficient broadband non-thermal emitter. In such a case, the created pairs can radiate through synchrotron and IC emission in
the whole spectral band, interacting with the ambient magnetic field and stellar photons, respectively \citep{bosch08a}. 

In case the energy density of the magnetic field is much smaller than that of the stellar photon field, IC becomes the
dominant cooling channel of these pairs. Then these pairs can produce more gamma-rays that to their turn may be also
absorbed, triggering what is called an electromagnetic cascade. In this way, the effective optical depth to gamma-rays can be
significantly reduced. The deflection of the created pairs in the ambient magnetic field determines whether the cascade is
linear or three dimensional \citep[e.g.][]{bednarek97,aharonian06b,orellana07,sierpowska07,zdziarski09,cerutti10}. 

\begin{figure} 
\center
\includegraphics[width=6.5cm]{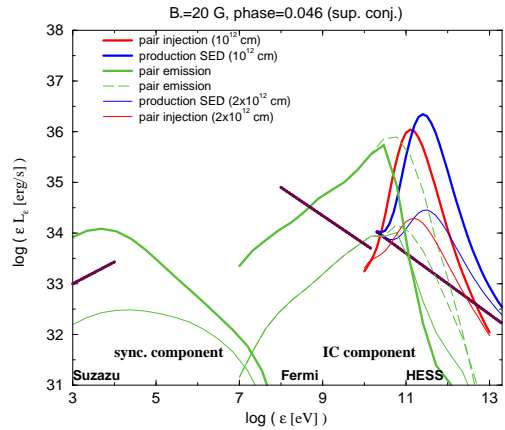}
\caption{\footnotesize
Computed spectral energy distribution of the radiation produced by secondary pairs in LS~5039 (green lines) at the orbital
phase associated to the superior conjunction of the compact object (the binary system properties can be found in 
Aragona et al. 2009). Two different configurations have been taken, one in
which the emitter is at a height of $10^{12}$~cm (production gamma-ray spectrum: thick solid blue line; corresponding pair
injection: thick solid red line), and at $2\times 10^{12}$~cm (production gamma-ray spectrum: thin solid blue line;
corresponding pair injection: thin solid red line) above the compact object. The production spectral energy distribution 
(thin long-dashed) of the secondary IC emission is also shown.
} 
\label{sec1}
\end{figure}

\begin{figure} 
\center
\includegraphics[width=6.5cm]{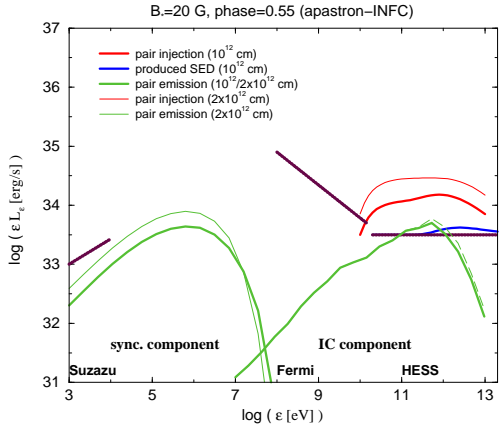}
\caption{\footnotesize
The same as in Fig.~\ref{sec1} but for the inferior conjunction of the compact object.
} 
\label{sec2}
\end{figure}

\begin{figure*}
\centering
\includegraphics[angle=0, width=0.4\textwidth]{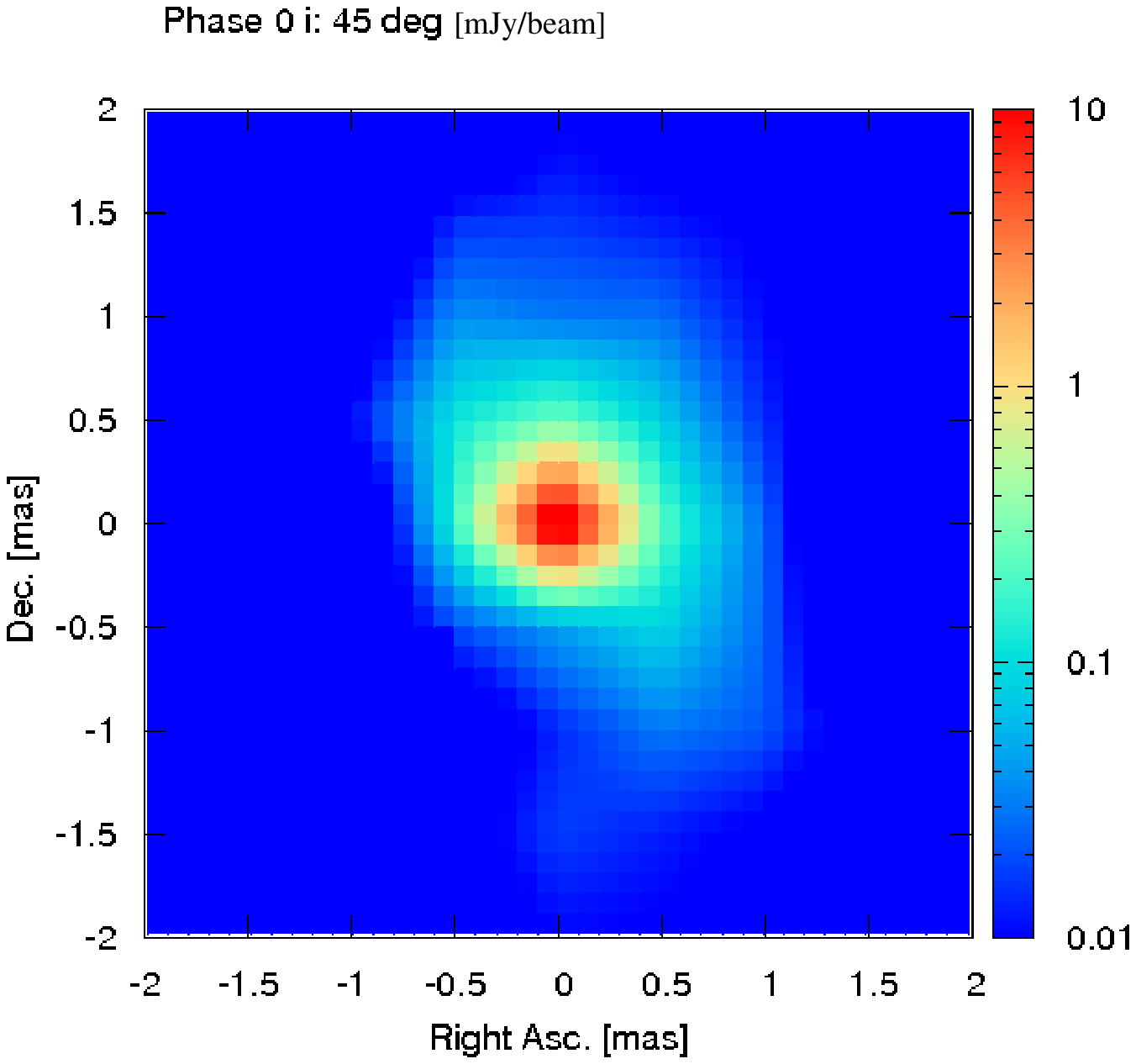}\qquad
\includegraphics[angle=0, width=0.4\textwidth]{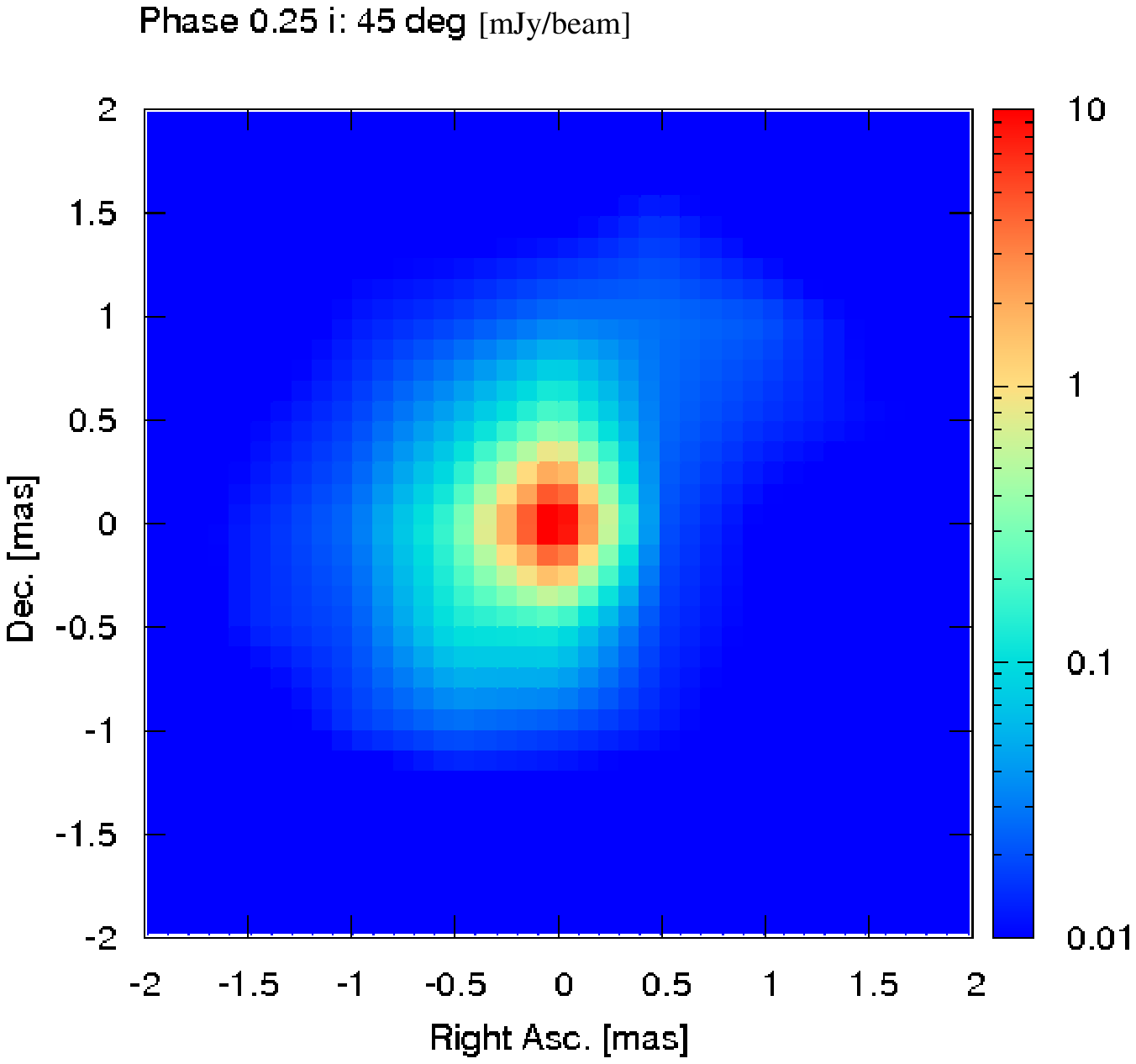}\\[10pt]
\includegraphics[angle=0, width=0.4\textwidth]{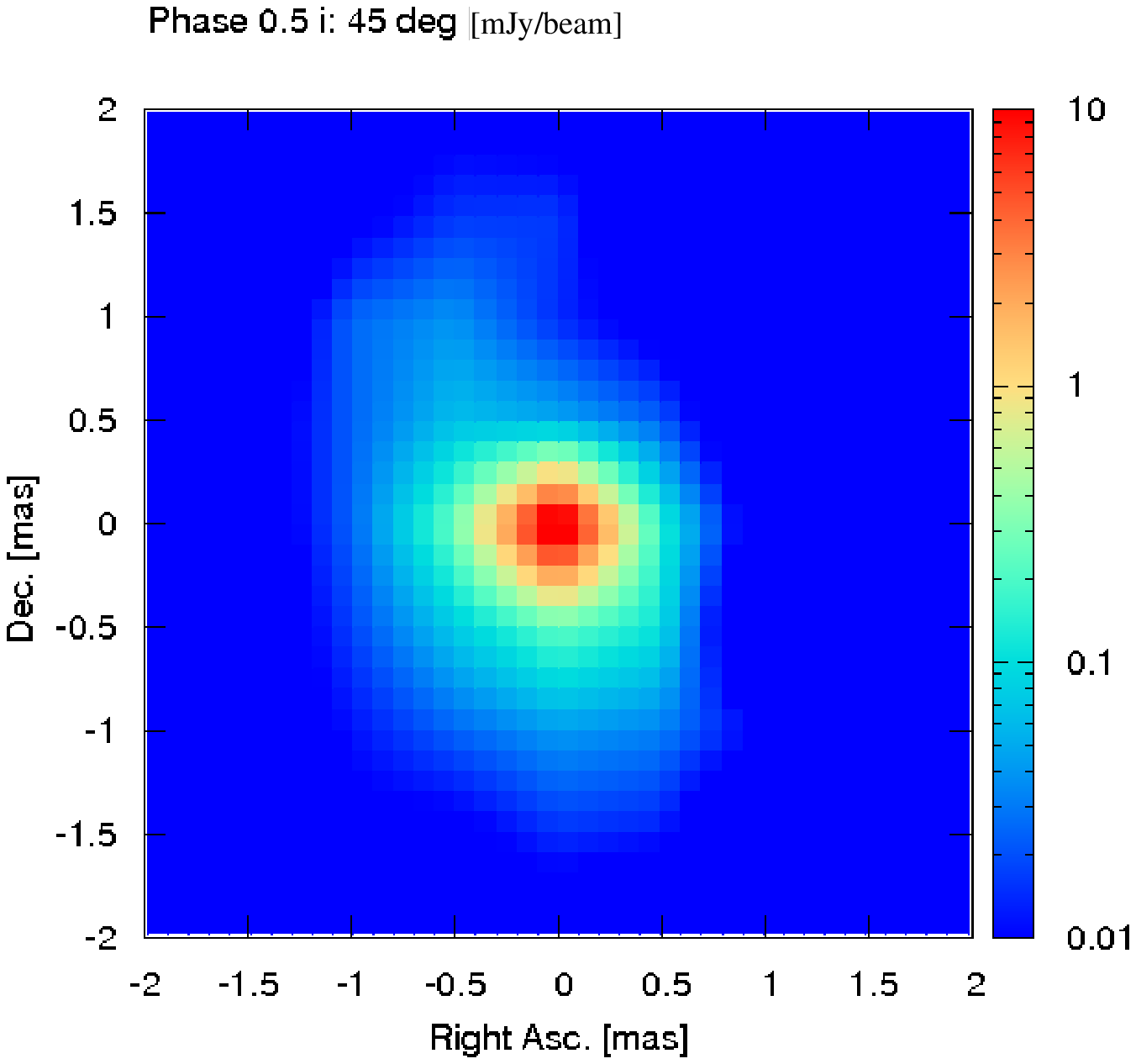}\qquad
\includegraphics[angle=0, width=0.4\textwidth]{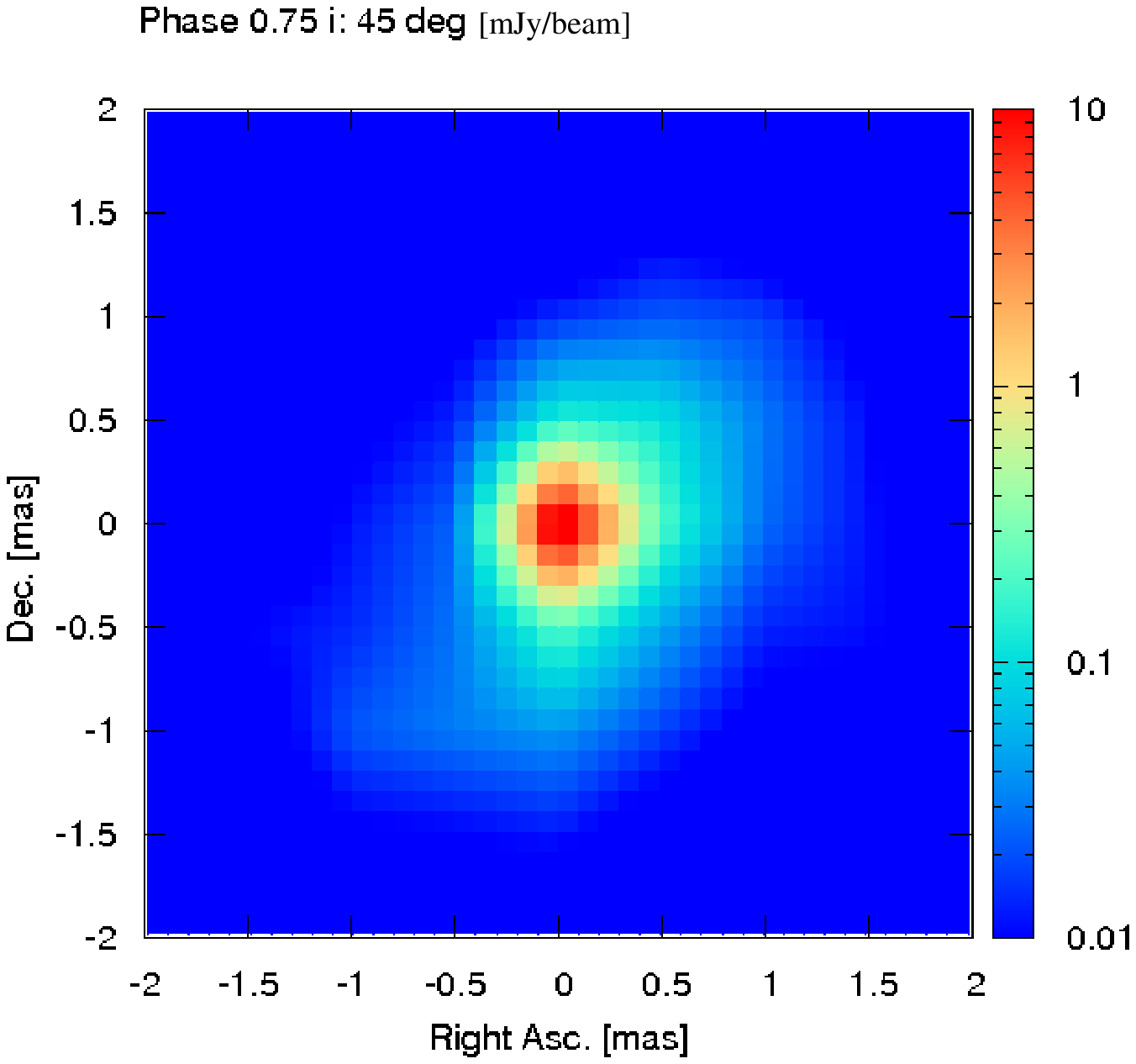}\\
\caption{Computed image of the 5~GHz radio emission, in the direction to the observer, for different orbital phases.
Units are given in mJy per beam, being the beam size $\sim 1$~milliarcsecond.}
\label{raw}
\end{figure*}

If the magnetic field is high enough, electromagnetic cascades are suppressed \citep{khangulyan08b} and the X-ray
(synchrotron) and GeV (single-scattering IC) emission from secondary pairs can dominate the secondary energy output
\citep{bosch08a}. As two examples of this, we show in Figs.~\ref{sec1} and \ref{sec2} the spectral energy distribution of the
secondary radiation for a gamma-ray binary with system properties similar to those of LS~5039. The magnetic field in the
stellar surface has been fixed to 20~G. For stronger enough magnetic fields, the secondary radio emission may be also
significant \citep{bosch09,bosch10b}. In Fig.~\ref{raw}, we show 5~GHz radio images for a similar case but adopting a star
magnetic field of 200~G. The images for four different orbital phases are presented: 0.0 (superior conjunction of the compact
object), 0.25, 0.5 (inferior conjunction), and 0.75. 

\section{Main elements to understand the gamma-ray emission}\label{fi}

The phenomenology at high energies of gamma-ray binaries can be (at least) semi-quantitatively explained accounting for a set
of ingredients or physical processes and effects: the angular dependence of gamma-ray absorption and IC scattering and
the (moderate) role of absorption \citep{khangulyan08b}, in some cases locating the emitter at some height above the orbital
plane (for acceleration and modeling arguments, see \citealt{khangulyan08b}; for absorption arguments, see
\citealt{bosch08b}), likely accounting for adiabatic losses \citep[e.g.][]{khangulyan07,takahashi09,zabalza10}.  The role of
the magnetic field, e.g. through the impact of synchrotron cooling, is crucial for the VHE spectrum and luminosity. In HMMQs,
one should not neglect the impact of the stellar wind on the non-thermal processes \citep{perucho10}.

All the mentioned physical effects and processes are present to different extents in all the known types of gamma-ray
binaries. This makes in some cases the identification of the specific nature of the object a difficult task, as it is shown
by the debate on the nature of LS~5039 and LS~I~+61~303, i.e. HMMQs versus non-accreting pulsar high-mass binaries, if not a
really new class of object. Therefore, it is of primary importance to apply comprehensive approaches to study theoretically
these objects, including semi-analytical modeling, numerical calculations (secondary emission) and (magneto)hydrodynamical
simulations, as well as carry out simultaneous multiwavelength observations with high angular, spectral and time resolution.
Nowadays, the field of gamma-ray binaries seems to be the best example for the need of such a complex multidisciplinary
approach, if one wants to understand the details of the physical processes taking place in these objects, as well as to
distinguish between different scenarios and models.

\begin{acknowledgements}
The author acknowledges support of the Spanish MICINN under grant
AYA2007-68034-C03-1 and FE\-DER funds, and thanks the Max Planck Institut 
f\"uer Kernphysik for its kind hospitality and support. The author also 
acknowledges support of
the European Community under a Marie Curie Intra-European fellowship.
\end{acknowledgements}

\bibliographystyle{aa}

\end{document}